\documentclass[tighten]{aastex701}
\usepackage{amssymb,amsmath,framed}
\usepackage{rotating,booktabs}
\usepackage{afterpage}
\usepackage{tikz}
\usepackage{tabularx}
\usepackage{savesym}
\savesymbol{tablenum}
\usepackage{siunitx}
\usepackage{caption}
\restoresymbol{SIX}{tablenum}
\usepackage{wrapfig}
\usepackage{hyperref}
\usepackage{bm}
\expandafter\ifx\csname package@font\endcsname\relax\else
 \expandafter\expandafter
 \expandafter\usepackage
 \expandafter\expandafter
 \expandafter{\csname package@font\endcsname}
\fi
\def\bq{\begin{equation}}
\def\eq{\end{equation}}
\def\bqy{\begin{eqnarray}}
\def\eqy{\end{eqnarray}}
\def\p{\partial}

\def\p{\partial}

\def\R{\mathbb{R}}

 \def\q#1#2{q^{\hspace{1pt}#1}_{\,,\hspace{1pt}#2}}
 \def\a#1#2{a^{\hspace{1pt}#1}_{\,,\hspace{1pt}#2}}
 \def\d#1#2{\delta^{\hspace{1pt}#1}_{\,,\hspace{1pt}#2}}

\begin{document}
 
\shorttitle{3I/ATLAS Solar Oberth Manoeuvre}
\shortauthors{Hibberd \& Eubanks}

\title{\large{Catching 3I/ATLAS Using a Solar Oberth}}

\correspondingauthor{Adam Hibberd}
\email{adam.hibberd@i4is.org}

\author[0000-0003-1116-576X]{Adam Hibberd}
\affiliation{Initiative for Interstellar Studies (i4is), 27/29 South Lambeth Road London, SW8 1SZ, United Kingdom}
\email{adam.hibberd@i4is.org}
\author[0000-0001-9543-0414]{T. Marshall Eubanks}
\affiliation{Space Initiatives Inc, Princeton, WV 24740, USA}
\email[show]{tme@space-initiatives.com}
\author[0000-0003-1763-6892]{Andreas M. Hein}
\affiliation{SnT, University of Luxembourg,  L-4365 Luxembourg}
\affiliation{Initiative for Interstellar Studies (i4is), 27/29 South Lambeth Road London, SW8 1SZ, United Kingdom}
\email{andreas.hein@i4is.org}

\begin{abstract}
The third interstellar object to be discovered, 3I/ATLAS, has a unique and continually unfolding story to tell of its nature and origin as it is monitored by telescopes on Earth, orbiting Earth and around the Solar System. Previous research into missions using chemical propulsion have only really addressed the direct case, where the opportunity to launch already expired before 3I/ATLAS's discovery. In contrast, investigations herein exploit 'Optimum Interplanetary Trajectory Software' to simulate an alternative indirect option for chemical propulsion, namely the Solar Oberth Manoeuvre (SOM). For a SOM, a low perihelion burn provides maximum benefit from the Oberth Effect, and accelerates the spacecraft rapidly towards the receding 3I/ATLAS. Though in principle feasible, results indicate this option presents significant challenges. For possible launch years between 2031 and 2037 inclusive, a 2035 launch permits the most efficient transfer to 3I/ATLAS. The reference mission requires a SOM at 3.2 Solar Radii from the Sun's centre, with an intercept after 35-50 years. It is found the SOM can leverage spacecraft masses up to $\sim{500}$ kg. Two or three solid propellant boosters could deliver the required SOM $\Delta$V, and furthermore a refuelled Starship Block 3 in LEO has sufficient performance for such a mission. As inevitable with a SOM, some of the payload mass would be needed for a heat shield to protect against the high solar flux at low perihelion.

\end{abstract}



\section{Introduction}
\label{sec1}

A spacecraft mission to 3I/ATLAS, the third interstellar object to be discovered passing through the Solar System \citep{seligman2025discovery,Loeb_2025,bolin2025,alvarezcandal2025,opitom2025,Chandler2025,Belyakov2025,hibberd2025interstellar}, presents major difficulties for a mission designer, particularly if a restriction of high TRL technologies is placed on the propulsion system, such as chemical rockets. Various researchers have investigated this case - \cite{Yaginuma_2025} delve into the potential for direct missions using chemical, whilst \cite{loeb2025intercepting} address the possibility of exploiting an existing spacecraft at Jupiter to achieve intercept.\\

The issues of a direct mission are centred largely on the celestial mechanics of the target in question and these are first the retrograde orientation of 3I/ATLAS's orbit, second the high heliocentric speed at infinity  of $\sim{60}$ $\si{km.s^{-1}}$ (otherwise known as the hyperbolic excess speed, V$_{\infty}$), and third the rather late initial detection, when 3I/ATLAS was already within the orbit of Jupiter.\\

The first of these effectively rules out a rendezvous mission, defined as one where the spacecraft uses on-board propulsion to match velocity with the target on arrival to allow prolonged close-up study of the body in-situ. Such a mission can be excluded because the high relative velocity between 3I/ATLAS and the spacecraft on arrival would be prohibitive. Thus only flyby missions are investigated here.\\

The second and third of these rule out a direct mission since the launch date of an optimal direct trajectory to 3I/ATLAS happens to be before its discovery. An ESA (European Space Agency) 'Comet Interceptor' type mission \citep{sanchez2025analysis} would seem to be a logical solution to this problem, since the architecture involves a spacecraft prepared and loitering in space, waiting to be dispatched to a suitable candidate almost immediately, with no accompanying launch delays. However, it can be shown that 3I/ATLAS would have required too high a thrust $\Delta$V even from a spacecraft already at the Sun/Earth Lagrange 2 point, ruling out the Comet Interceptor strategy, at least in its current form.\\

To exacerbate these three aforementioned impediments, there is now the additional complication that 3I/ATLAS has flown past its perihelion (which was on 2025 October 29$^{th}$) and is now rapidly receding the Sun at a speed in excess of 60 \si{km.s^{-1}}. This necessitates any future mission catching-up with the target, and so requiring a heliocentric V$_{\infty}$ far in excess of this value to stand any chance of paying a visit.\\

It is for these above reasons that one might consider the so-called 'Solar Oberth Manoeuvre' (SOM) option \citep{oberth2019wege} as one of the few remaining candidates, an architecture researched extensively and exhaustively by Project Lyra \citep{HPE19,HEL22,HHE20,HH21,AH23,HA23}, the feasibility study instigated by the 'Initiative for Interstellar Studies' \citep{I4IS} into missions to the first discovered interstellar object (ISO) 1I/'Oumuamua . Also refer to \cite{Seligman_2018} who studied direct missions to 1I/'Oumuamua.\\

Similar SOM research was conducted by the i4is team on 2I/Borisov \citep{HPH21}.

\section{Method}

To solve the optimal trajectories the preliminary interplanetary mission design tool known as 'Optimum Interplanetary Trajectory Software' (OITS) was deployed \citep{OITS_info,AH2,HPH21}, conceived, developed, validated and applied by Adam Hibberd, and used for most of the Project Lyra research.\\

Let's say we have $N$ encounters of celestial bodies along a trajectory, where the $N$ bodies include the initial home planet (Earth) as well as the final target celestial body (in this case 3I/ATLAS). We associate with each encounter a mission time, and it is this sequence of times which must be 'optimized', in the sense of minimizing the overall trajectory $\Delta$V.\\

But how is this total $\Delta$V calculated from the encounter times, $t_i, i=1,2...N$ ?\\

First, for each pair of consecutive bodies ($i, i+1$), we have their positions ($r_i, r_{i+1}$) and velocities ($v_i, v_{i+1}$) at their respective encounter times (there are innumerable ways to calculate these, it is sufficient here to know that this can be done). The problem of solving an orbit to connect $r_i$ to $r_{i+1}$ in an elapsed time ($t_{i+1}-t_i$), is known as the Lambert Problem and is solved by OITS using the well-established 'Universal Variable' formulation \citep{Bate1971}.\\

Having so-derived these $N-1$ connecting orbits, we can denote the arrival velocity
 relative to body $i$ as $VA_i$ and the departure velocity $VD_i$, relative to $i$, which is the velocity required to reach body $i+1$. A 'patched conic' simplification is then applied whereby these are assumed to be the hyperbolic excess speeds of arrival and exit relative to the encounter body in question ($i$). If we further assume that we wish to reap maximum benefit from the 'Oberth Effect', then it follows the spacecraft must apply all its $\Delta$V at periapsis w.r.t. the body in question. Furthermore there turns out to be only one combination of periapsis altitude and $\Delta$V capable of satisfying the ingoing and outgoing velocities $VA_i$ and $VD_i$. This allows solution of the $\Delta$V problem. Alternatively, if the encounter body has no significant mass, then the $\Delta$V is simply given by $VD_i-VA_i$.\\

 It is now a case of summing all these individual $\Delta$V$_i$ to get the total $\Delta$V for the trajectory in question. A Non-Linear Problem Solver is then utilized to find the combination $t_i, i=1,2...N$ which minimizes total $\Delta$V. Moreover, extra inequality constraints need to be introduced to ensure each calculated periapsis altitude is $ > 200~ \si{km}$, and additional $\Delta$V$_i$ restrictions for each encounter may be enforced.\\
 
 The challenge of modelling the SOM is one of finding a point in space which has NO mass and so whose $\Delta$V is $VD_i-VA_i$, as indicated. To this end the notion of the 'Intermediate Point' \citep{AH2} can be introduced, this being a point whose Sun-distance, $R_S$, is user-specified, but whose heliocentric longitude and latitude, $\theta$ and $\phi$ are unknown and need to be optimized so as to minimize overall $\Delta$V. Thus these are thrown in with the previously mentioned encounter times and presented to the NLP for optimization.\\

 Finally the position and velocities of ALL celestial bodies as a function of time (alluded to above) are found using the NASA JPL NAIF SPICE toolkit \citep{NAIF} whereby NASA SPICE kernels for all encounter bodies can be downloaded from the website and exploited for this purpose. Two NLP solvers are utilised for this research, NOMAD \citep{LeDigabel2011} and MIDACO \citep{Schlueter_et_al_2009,Schlueter_Gerdts_2010,Schlueter_et_al_2013}.\\

 We now discuss the trajectory scenario investigated in this paper. This can be written as E-J-SOM-3I, including home and target bodies. Note that there is a Jupiter encounter (J) after the Earth launch (E) and before the SOM. This Jupiter encounter is included for good reason since it allows leverage of Jupiter's huge mass to slow down the spacecraft, in turn by effectively nullifying its tangential velocity (endowed by Earth's heliocentric velocity). It is only by using this device that a low perihelion can be achieved by a chemically propelled spacecraft, since to lose all of Earth's velocity of $\sim{30}$ $\si{km.s^{-1}}$ otherwise would be extremely challenging.\\

 In addition, a constraint is placed on Jupiter $\Delta$V to force it to a negligible level (for various practical reasons there may be a small residual velocity increment at Jupiter but this can effectively be discounted), thus leading to passive Jupiter gravitational assists (GAs) and greatly simplifying the mission architecture so that all the on-board rocketry can be dedicated to the SOM when it happens.\\

\section{Results}
 An earliest launch date 5 years hence (i.e. in the year 2031) and up to 11 years in the future is assumed, equivalent to the launch years 2031 to 2037. The situation is summarized in the 4 Figures provided.\\
\begin{figure}[hbt!]
\centering
\includegraphics[width=0.9\textwidth]{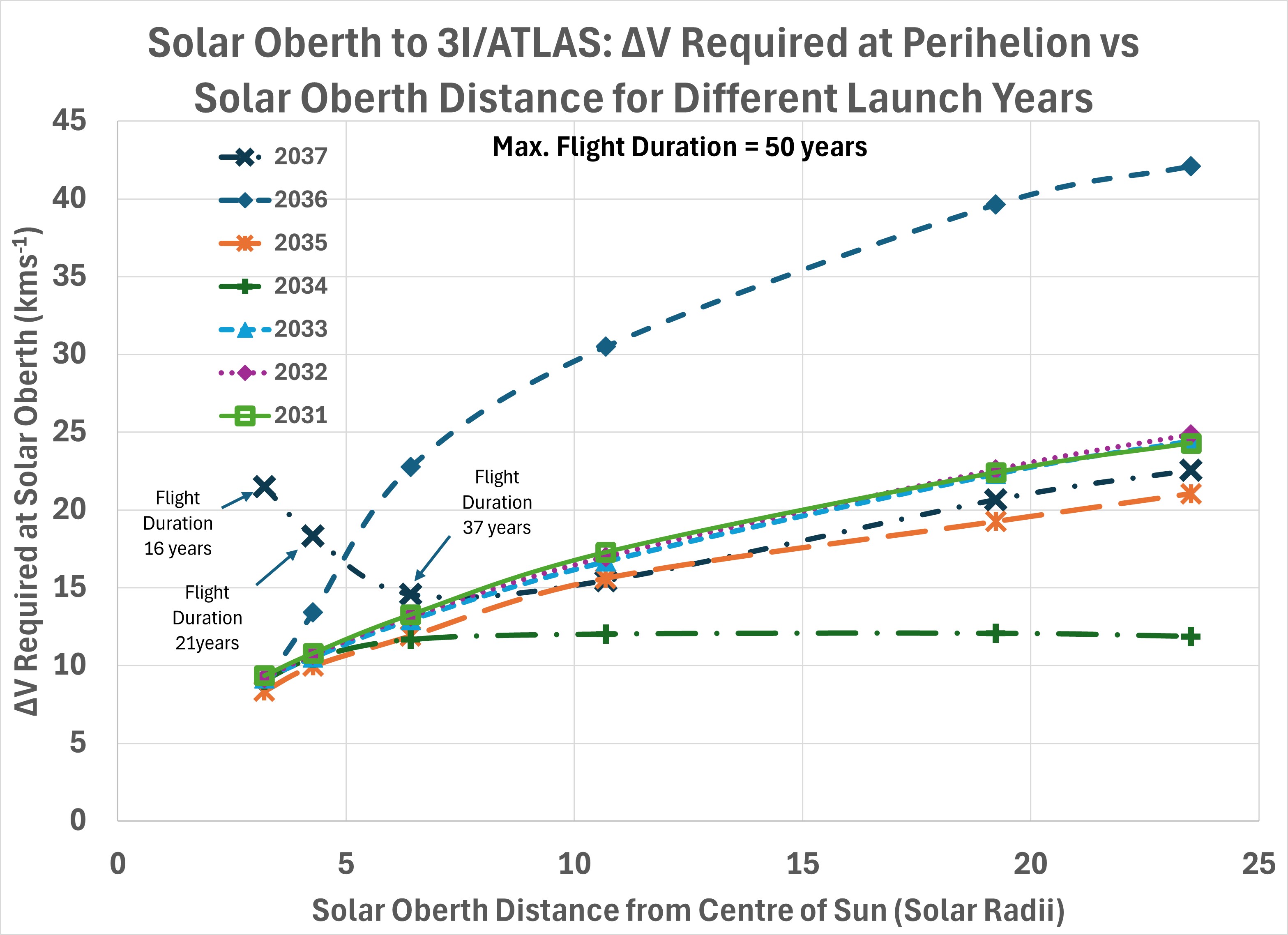}
\caption{$\Delta$V needed at SOM against perihelion distance for missions with E-J-SOM-3I sequence and assuming no $\Delta$V at Jupiter.}
\label{fig:SOMDELTAV}
\end{figure}

 The first of these (Figure \ref{fig:SOMDELTAV}) provides the necessary $\Delta$V needed at the SOM to achieve an intercept of 3I/ATLAS within the span of 50 years from launch. The minimum value is attained for a launch in 2035 and corresponds to the lowest Solar Oberth distance investigated here, namely 0.015 au, equivalent to 3.2 Solar Radii (SR) from the Sun's centre. This minimum $\Delta$V is 8.36 $\si{km.s^{-1}}$. It can be compared with the default Solar Oberth mission plan studied by Project Lyra (for 1I/'Oumuamua) which had a SOM $\Delta$V of 7.2 $\si{km.s^{-1}}$, a perihelion of 6SR and a mission duration of 22 years.\\
 
\begin{table}[]
\centering
\caption{Table of possible mission flight durations for a launch year 2035 and with a SOM burn at 3.2SR (Solar Radii) from the centre of the Sun \label{tab:2035}}
\begin{tabular}{lllllllllll}
\cline{1-8}
\multicolumn{1}{|c|}{\textbf{}} &
  \multicolumn{1}{c|}{\textbf{}} &
  \multicolumn{1}{c|}{\textbf{}} &
  \multicolumn{1}{c|}{\textbf{}} &
  \multicolumn{1}{c|}{\textbf{Maximum}} &
  \multicolumn{1}{c|}{\textbf{Arrival}} &
  \multicolumn{1}{c|}{\textbf{CASTOR 30B}} &
  \multicolumn{1}{c|}{\textbf{CASTOR 30B}} &
   &
   &
   \\
\multicolumn{1}{|c|}{\textbf{Flight}} &
  \multicolumn{1}{c|}{\textbf{Intercept}} &
  \multicolumn{1}{c|}{\textbf{C$_3$}} &
  \multicolumn{1}{c|}{\textbf{SOM $\Delta$V}} &
  \multicolumn{1}{c|}{\textbf{Heliocentric}} &
  \multicolumn{1}{c|}{\textbf{Speed Rel.}} &
  \multicolumn{1}{c|}{\textbf{STAR 48B}} &
  \multicolumn{1}{c|}{\textbf{STAR 48B}} &
   &
   &
   \\
\multicolumn{1}{|c|}{\textbf{Duration}} &
  \multicolumn{1}{c|}{\textbf{Distance}} &
  \multicolumn{1}{c|}{\textbf{}} &
  \multicolumn{1}{c|}{\textbf{}} &
  \multicolumn{1}{c|}{\textbf{speed at SOM}} &
  \multicolumn{1}{c|}{\textbf{to 3I/ATLAS}} &
  \multicolumn{1}{c|}{\textbf{Payload}} &
  \multicolumn{1}{c|}{\textbf{Total}} &
   &
   &
   \\
\multicolumn{1}{|c|}{\textbf{($yrs$)}} &
  \multicolumn{1}{c|}{\textbf{($au$)}} &
  \multicolumn{1}{c|}{\textbf{($km^2s^{-2}$)}} &
  \multicolumn{1}{c|}{\textbf{($kms^{-1}$)}} &
  \multicolumn{1}{c|}{\textbf{($kms^{-1}$)}} &
  \multicolumn{1}{c|}{\textbf{($kms^{-1}$)}} &
  \multicolumn{1}{c|}{\textbf{($kg$)}} &
  \multicolumn{1}{c|}{\textbf{($kg$)}} &
   &
   &
   \\ \cline{1-8}
\multicolumn{1}{|c|}{50} &
  \multicolumn{1}{c|}{732} &
  \multicolumn{1}{c|}{130.2} &
  \multicolumn{1}{c|}{8.355} &
  \multicolumn{1}{c|}{352} &
  \multicolumn{1}{c|}{16} &
  \multicolumn{1}{c|}{546} &
  \multicolumn{1}{c|}{17754} &
   &
   &
   \\ \cline{1-8}
\multicolumn{1}{|c|}{40} &
  \multicolumn{1}{c|}{609} &
  \multicolumn{1}{c|}{130.1} &
  \multicolumn{1}{c|}{9.291} &
  \multicolumn{1}{c|}{353} &
  \multicolumn{1}{c|}{20} &
  \multicolumn{1}{c|}{342} &
  \multicolumn{1}{c|}{16450} &
   &
   &
   \\ \cline{1-8}
\multicolumn{1}{|c|}{30} &
  \multicolumn{1}{c|}{487} &
  \multicolumn{1}{c|}{162.1} &
  \multicolumn{1}{c|}{10.36} &
  \multicolumn{1}{c|}{354} &
  \multicolumn{1}{c|}{25} &
  \multicolumn{1}{c|}{N/A} &
  \multicolumn{1}{c|}{N/A} &
   &
   &
   \\ \cline{1-8}
\multicolumn{1}{|c|}{20} &
  \multicolumn{1}{c|}{365} &
  \multicolumn{1}{c|}{175.4} &
  \multicolumn{1}{c|}{14.077} &
  \multicolumn{1}{c|}{357} &
  \multicolumn{1}{c|}{39} &
  \multicolumn{1}{c|}{N/A} &
  \multicolumn{1}{c|}{N/A} &
   &
   &
   \\ \cline{1-8}
\multicolumn{1}{|c|}{10} &
  \multicolumn{1}{c|}{239} &
  \multicolumn{1}{c|}{955.1} &
  \multicolumn{1}{c|}{29.991} &
  \multicolumn{1}{c|}{372} &
  \multicolumn{1}{c|}{85} &
  \multicolumn{1}{c|}{N/A} &
  \multicolumn{1}{c|}{N/A} &
   &
   &
   \\ \cline{1-8}
\end{tabular}
\end{table}

 It is quite clear from this that, largely due to the 3 factors elucidated in the introduction, 3I/ATLAS presents a far more challenging prospect than that of 1I/'Oumuamua for a mission to an ISO. \\
 
 Comparing 2035 with other launch years, we observe virtually identical  $\Delta$V profiles for 2031, 2032 and 2033. However we find that launch year 2037 has a similar curve to 2031/2032/2033 and 2035, except where the 2037 perihelia drop below around 6SR at which point the $\Delta$V begins to escalate in comparison, though with the benefit of reduced flight durations, as marked on the plot.\\

 Missions launching in 2034 and 2036 have respectively, lower and higher SOM $\Delta$Vs than all the other options investigated thus it seems sensible to reject 2036 as a launch year, save for the lowest extreme of perihelion - i.e. the 3.2SR case.\\
 
\begin{figure}[hbt!]
\centering
\includegraphics[width=0.9\textwidth]{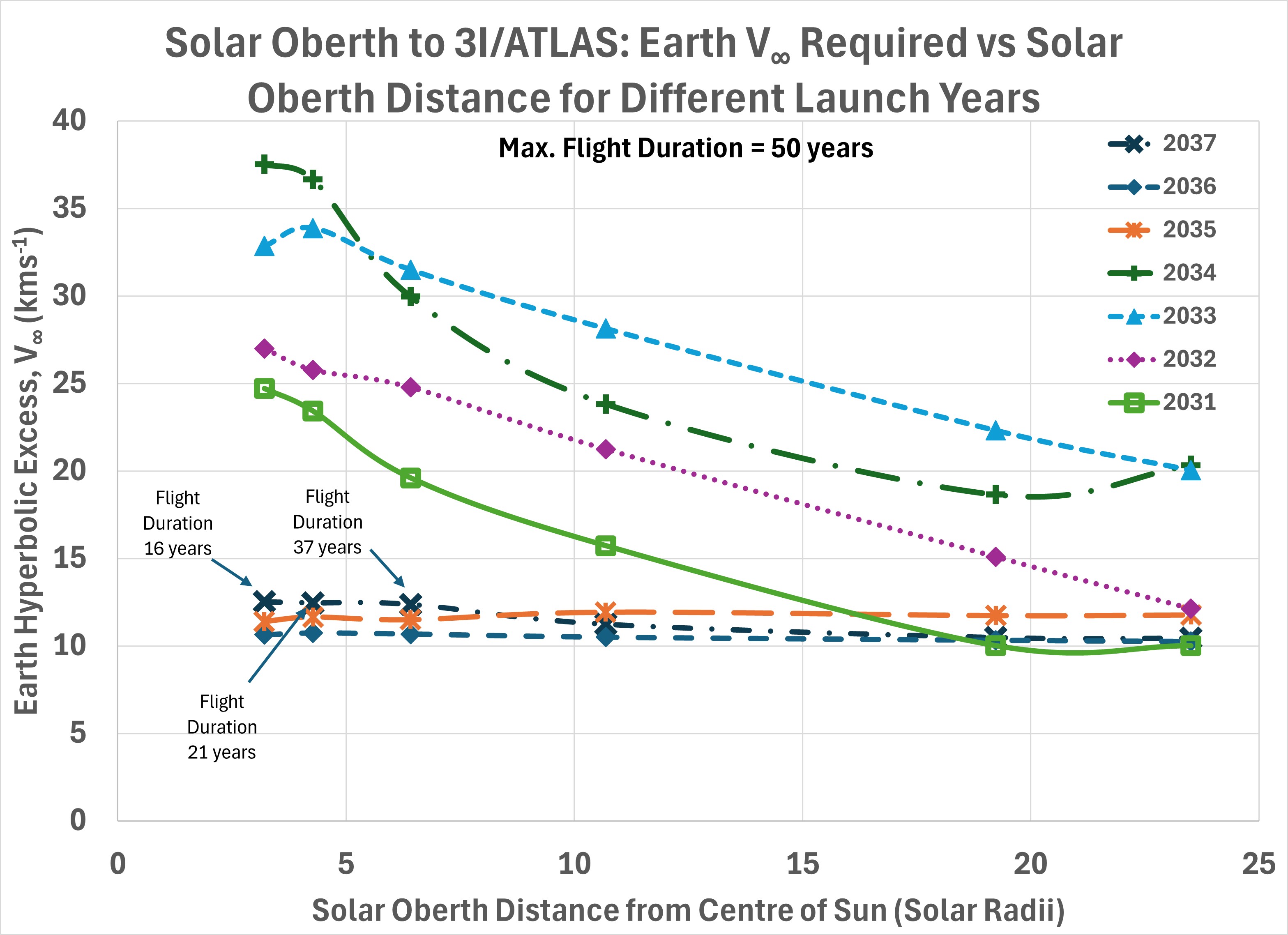}
\caption{Earth hyperbolic excess speed, $V_{\infty}$ against perihelion distance needed for missions with the E-J-SOM-3I sequence and assuming no $\Delta$V at Jupiter.}
\label{fig:SOMVINF}
\end{figure}
 We now look into the required geocentric hyperbolic excess ($V_{\infty}$) needed by the launch vehicle to reach Jupiter and so set up the conditions for the SOM. Ideally, this should be minimized to maximize the mission spacecraft payload mass. To this end refer to Figure \ref{fig:SOMVINF}, which shows precisely this metric for each launch year and for each perihelion distance in turn.\\

 We find that from years 2035 to 2037 inclusive, the level of $V_{\infty}$ stays low, at around $\sim{13}$
$\si{km.s^{-1}}$, and all other options (2031-2034) have much more demanding values, especially at low perihelia, making these missions largely untenable practically speaking. The launch year 2036 can be rejected on grounds of too high a SOM $\Delta$V, leaving it unfavourable compared to the alternatives.\\

\begin{figure}[hbt!]
\centering
\includegraphics[width=0.9\textwidth]{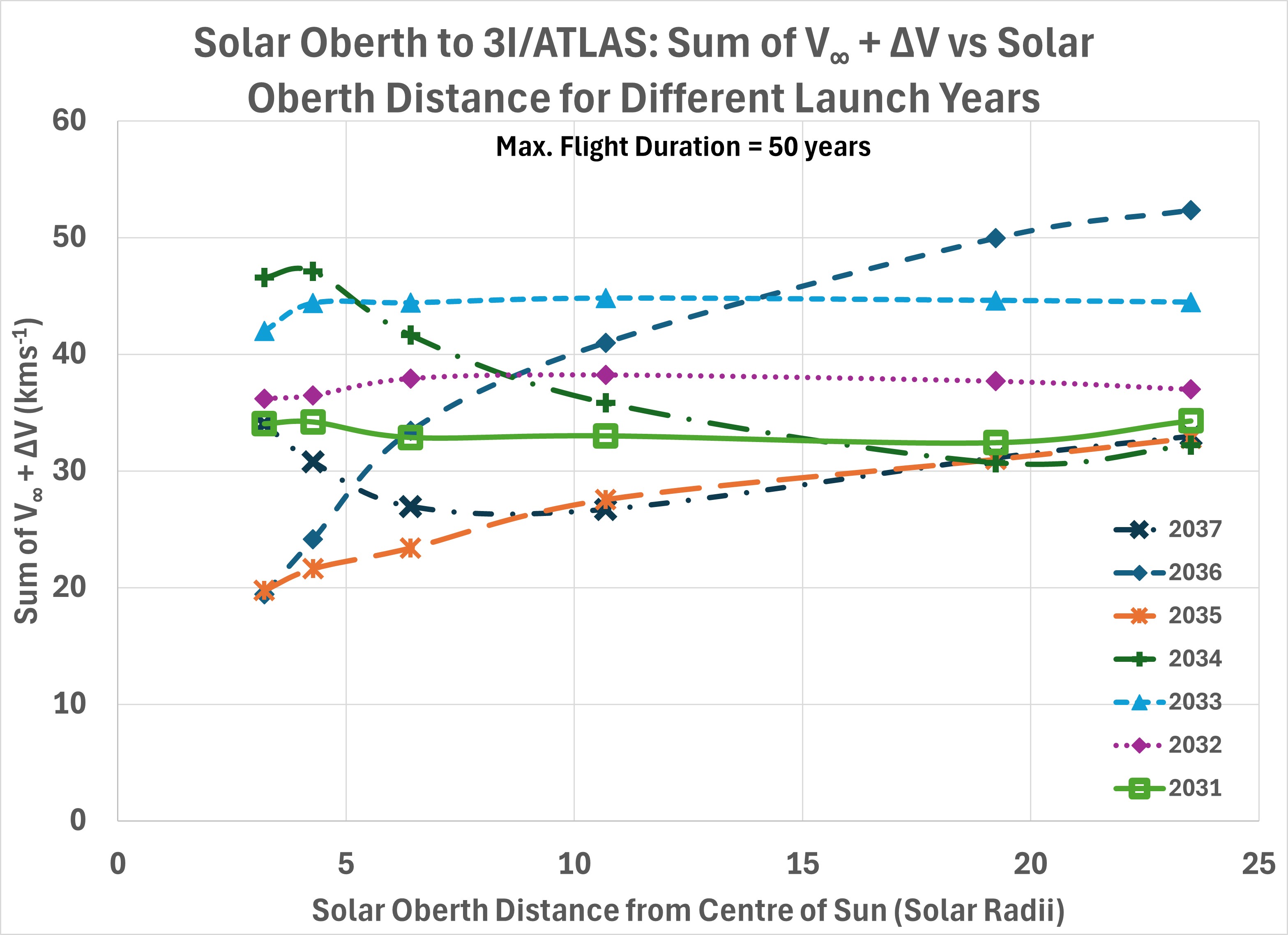}
\caption{Sum of Figures \ref{fig:SOMDELTAV} and \ref{fig:SOMVINF}}
\label{fig:TOTDELTAV}
\end{figure}
The general feasibility of missions to 3I/ATLAS is summarized in Figure \ref{fig:TOTDELTAV} which does not actually represent any real physical parameter, it is simply the sum of the SOM $\Delta$V and the Earth $V_{\infty}$, provided in Figures \ref{fig:SOMDELTAV} and \ref{fig:SOMVINF} respectively, from which we can grasp at the general feasibility of missions for the launch years studied.\\

To this end we find that 2035 is the most propitious year for launch to 3I/ATLAS, as it out performs all other launch years except for 2036 (but only at 3.2SR). Also observe that 2037 presents an additional decent opportunity, with the higher requirements on the SOM $\Delta$V being mitigated by much lower flight durations, eg 21 years for 4.3SR perihelion or 16 years for the 3.2SR perihelion.\\

\begin{figure}[hbt!]
\centering
\includegraphics[width=0.9\textwidth]{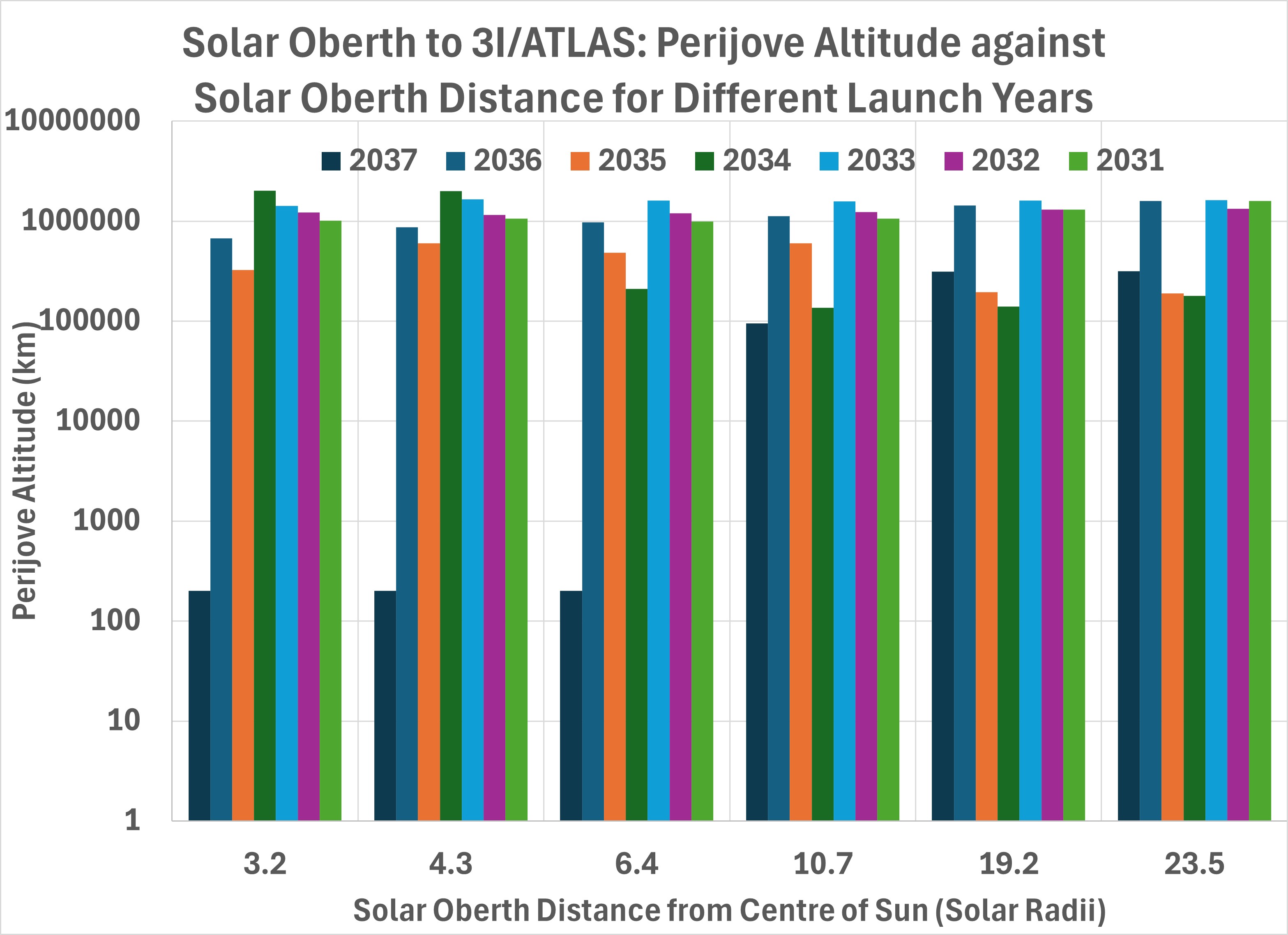}
\caption{Perijove altitudes on a logarithmic scale for launch years 2031-2037 and for different Solar Oberth perihelia.}
\label{fig:PERIJOVE}
\end{figure}

To get an idea of the level of perispasis altitude at the Jupiter encounters, otherwise known as 'perijove', refer to Figure \ref{fig:PERIJOVE}. We find it is generally and comfortably in the region 100,000 $\si{km}$ to 2,000,000 $\si{km}$.\\

Since the 2035/3.2SR combination appears to offer optimal performance for the SOM strategy with flight duration constraint of 50 years, it would be instructive to determine how this trajectory is affected by reducing flight duration - thus refer to Table \ref{tab:2035}. We find that missions become untenable below a flight duration of between 30 and 40 years for this trajectory, the relevance of the masses provided on the right of this table is articulated in the next Section.

\section{Discussion}

It is not intended here to embark on a detailed mission analysis, with a mass budget, $\Delta$V budget, scientific instrumentation, etc. The research herein is intended as a general feasibility study for a Solar Oberth to 3I/ATLAS. It has been shown here that such a mission is extremely challenging, even for the optimal launch year of 2035. But for the moment let us adopt this year as a reference.\\

Reference to Figure \ref{fig:SOMDELTAV} reveals that the overwhelming requirement of such a mission is the delivery of a SOM $\Delta$V of 8.36 $\si{km.s^{-1}}$ at 3.2SR. Is such a high $\Delta$V even practical? Let us now examine the chemical propulsion alternatives - in fact solid rocket stages - we have available to us. This analysis is summarized in Table \ref{tab:SOM}. Table \ref{tab:SOM} details different permutations involving either 3 stages or 2 stages of solid propellant rockets, drawn from a list of 5 commerically available off-the-shelf solid boosters. These are in order of decreasing performance, CASTOR 30XL, CASTOR 30B, STAR 75, STAR 63F and STAR 48. Further provided in this table are several example combinations, all of which permit a positive payload mass to be delivered by the end of the SOM burn. Note this is not an exhaustive list of all options.\\

\afterpage{
\begin{sidewaystable*}[]

{\small\setlength{\tabcolsep}{1pt} 
\captionsetup{justification=centering}
\caption{\label{tab:SOM}Summary of different solid rocket stage combinations enabling an overall SOM $\Delta$V of 8.36 $\si{km.s^{-1}}$, assuming first 3 and then 2 stages. Note the payload mass and also the total mass of the spacecraft are provided for each combination studied. The scenarios achievable by a fully refuelled Starship Block 3 in LEO are ticked, representing feasible missions.}
\hspace{-5.0cm}
\vspace{1.0cm}
\begin{tabular}{|c|c|c|c|c|c|c|c|c|c|l|}
\hline
\multicolumn{1}{|l|}{\textbf{}} &
  \textbf{STAR 75} &
  \textbf{STAR 63F} &
  \textbf{STAR 48} &
  \textbf{CASTOR 30XL} &
  \textbf{CASTOR 30B} &
  \textbf{} &
  \textbf{} &
  \multicolumn{1}{l|}{\textbf{}} &
  \multicolumn{1}{l|}{\textbf{}} &
   \\ \hline
\multicolumn{1}{|l|}{\textbf{Tot. Mass (kg)}} &
  8068 &
  4590 &
  2137 &
  26406 &
  13970.6 &
   &
   &
  \multicolumn{1}{l|}{} &
  \multicolumn{1}{l|}{} &
   \\ \hline
\multicolumn{1}{|l|}{\textbf{Dry Mass (kg)}} &
  565 &
  326 &
  124 &
  1392 &
  1000 &
   &
   &
  \multicolumn{1}{l|}{} &
  \multicolumn{1}{l|}{} &
   \\ \hline
\multicolumn{1}{|l|}{\textbf{Exh. Vel. (km/s)}} &
  2.8224 &
  2.9106 &
  2.8028 &
  2.8866 &
  2.9649 &
   &
   &
   &
  \textbf{Payload} &
  \multicolumn{1}{c|}{\textbf{Starship}} \\ \hline
\multicolumn{1}{|l|}{\textbf{Length (m)}} &
  2.591 &
  2.71 &
  2.075 &
  5.99 &
  4.315 &
  \textbf{Payload Mass (kg)} &
  \textbf{Total Mass (kg)} &
  \textbf{Length (m)} &
  \textbf{Mass Frac(\%)} &
  \multicolumn{1}{c|}{\textbf{Capability?}} \\ \hline
\multicolumn{1}{|l|}{\textbf{3 Stages:}} &   &   &     &   &   &       &       &       & \multicolumn{1}{l|}{} &                        \\ \hline
\textbf{a}                               & 3 &   &     & 1 & 2 & 1790 & 50234 & 12.90 & 3.56                 &                        \\ \hline
\textbf{b}                               & 2 &   & 3   & 1 &   & 1420 & 38031 & 10.66 & 3.73                 &                        \\ \hline
\textbf{c}                               &   & 2 & 3   & 1 &   & 1332 & 34465 & 10.78 & 3.86                 &                        \\ \hline
\textbf{d}                               & 2 &   & 3   &   & 1 & 909   & 25085 & 8.98  & 3.62                  &                        \\ \hline
\textbf{e}                               &   & 2 & 3   &   & 1 & 820   & 21518 & 9.10  & 3.81                  &                        \\ \hline
\textbf{f}                               &   & 1 & 2+3 &   &   & 449   & 21175 & 6.86  & 2.12                  &                        \\ \hline
\textbf{g}                               & 1 & 2 & 3   &   &   & 544   & 15339 & 7.38  & 3.55                  & \multicolumn{1}{c|}\checkmark \\ \hline
\textbf{h}                               & 1 &   & 2+3 &   &   & 430   & 12772 & 6.74  & 3.37                  & \multicolumn{1}{c|}\checkmark \\ \hline
\multicolumn{1}{|l|}{\textbf{2 Stages:}} &   &   &     &   &   &       &       &       &                       &                        \\ \hline
\textbf{i}                               &   &   &     & 1 & 2 & 1217  & 41593 & 10.31 & 2.93                  &                        \\ \hline
\textbf{j}                               & 2 &   &     & 1 &   & 1075  & 35549 & 8.58  & 3.02                  &                        \\ \hline
\textbf{k}                               & 2 &   &     &   & 1 & 526   & 22565 & 6.91  & 2.33                  &                        \\ \hline
\textbf{l}                               &   & 2 &     &   & 1 & 602   & 19163 & 7.03  & 3.14                  &                        \\ \hline
\textbf{m}                               &   &   & 2   &   & 1 & 546   & 17754 & 6.39  & 3.08                  & \multicolumn{1}{c|}\checkmark \\ \hline
\textbf{n}                               & 1 & 2 &     &   &   & 315   & 12973 & 5.30  & 2.43                  & \multicolumn{1}{c|}\checkmark \\ \hline
\textbf{o}                               & 1 &   & 2   &   &   & 312   & 10518 & 4.67  & 2.97                  & \multicolumn{1}{c|}\checkmark \\ \hline
\end{tabular}}

\end{sidewaystable*}
}

This reveals that delivery of such a high SOM $\Delta$V is feasible, but would there be a launch vehicle capable of inserting any of these combinations into the required Earth-escape orbit?\\

The Earth hyperbolic target for the 2035 scenario we are addressing has a launch characteristic energy (C$_3$ - defined as $V_{\infty}^2$) of 130.2 $\si{km^2.s^{-2}}$. If we now assume that the SpaceX Starship Block 3 will be operating routinely by this time, and further that the long-term intention for this launch vehicle, i.e. to allow LEO (Low Earth Orbit) accessibility for the upper stage at very low cost, has been accomplished, then the plan is that this should enable refuelling of the Starship upper stage to be conducted which in turn permits the orbiting Starship to inject into escape orbits to the planets, like a Mars destination for instance, but alternatively and especially applicable here, to the planet Jupiter.\\

\begin{figure}[hbt!]
\centering
\includegraphics[width=0.9\textwidth]{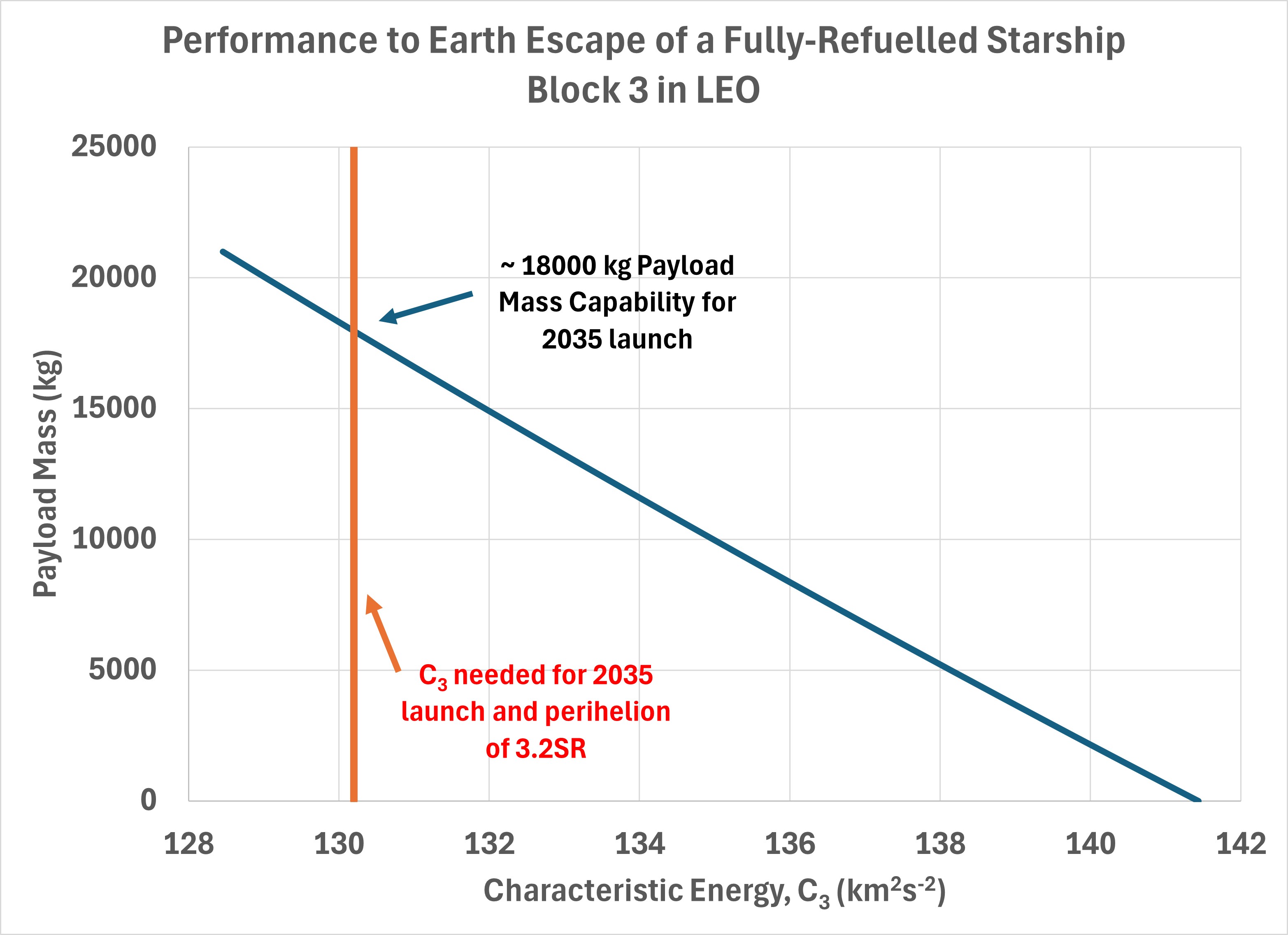}
\caption{Performance capability of a Starship Block 3 fully refuelled and in LEO in terms of total payload mass (including additional SOM stages) against Characteristic Energy, $C_3$}
\label{fig:STARSHIP}
\end{figure}

Now refer to Figure \ref{fig:STARSHIP} and we find that for the aforementioned required C$_3$, a total payload mass of 18000 $\si{kg}$ can be accommodated.\\

Referring back to Table \ref{tab:SOM}, there are 5 scenarios which could be lofted by the Starship since their total mass turns out to be $<$ 18000 $\si{kg}$, as required. The scientific payload delivered for these 5 scenarios range from 312 $\si{kg}$ to 546 $\si{kg}$, as a comparison the New Horizons spacecraft to Pluto was around 500 $\si{kg}$.\\

Note that the solar flux at 3.2 au from the Sun's centre is extremely high, i.e. $\sim{6}$ $\si{MWm^{-2}}$ so a heat shield, such as proposed in \cite{benkoski2019physics} would need to be included in the mass budget and would detrimentally impact on the useful payload mass arriving at 3I/ATLAS.\\

One may contemplate in the light of these investigations whether a mission to 3I/ATLAS is worth all the trouble, especially in view of the long mission durations envisioned of 35-50 years with modern day technology. The case for a mission is tempered by the wealth of data already garnered by Earth, space and deep space telescopes across the Solar System, and this over a considerable period of time. As a result of this astronomical curiosity, we have a good knowledge of its composition, and 3I/ATLAS has not exactly hidden its secrets from us as was the case for 1I/'Oumuamua. On the contrary, 3I/ATLAS is a very unusual comet of a unique kind, and as a comet has exhibited invaluable spectral patterns allowing us to sneak an insight into its origins and nature.\\

However, like so many celestial curiosities which abide in our own Solar System for example, there is nonetheless a lot to gain from a close encounter with this object by a spacecraft, and furthermore myriad questions, otherwise lost to the realms of conjecture, supposition or speculation, could only be answered properly once the cometary display has faded, by a spacecraft mission. Also bear in mind that the level of encounter velocity for a 50 year mission (ref. Table \ref{tab:2035}) is quite low at 16 $\si{km.s^{-1}}$ compared to the nominal 1I/'Oumuamua mission (30 $\si{km.s^{-1}}$).\\

\begin{table}[h!]
\centering
\caption{Approximate Comparative Positional Uncertainties at 100 AU}
\label{tab:errors}
\begin{tabular}{lcc}
\toprule
Parameter & 1I/'Oumuamua & 3I/ATLAS \\
\midrule
Transverse Error (km) & 200,000 & 2,000 \\
Radial Error (km)     & 1,000,000 & 20,000 \\
\bottomrule
\end{tabular}
\end{table}

To perform a flyby of an interstellar object in deep space, its position needs to be known with sufficient accuracy. We argue that for 3I/ATLAS, the position will be known with sufficient accuracy for a deep space flyby. This is in contrast to the first interstellar visitor, 1I/'Oumuamua, whose non-gravitational accelerations and dim nature would make a flyby particularly challenging \citep{hein2022interstellar}. \\

3I/ATLAS's radial velocity error is, right now, 80 $\si{mm.s^{-1}}$, even before most of the spacecraft astrometry is in hand. That's about a factor of 20 times better than 1I/'Oumuamua in the radial direction. In the transverse, 3I/ATLAS is known $\sim{100}$ times better than 1I/'Oumuamua in terms of angular units. Since its V$_{\infty}$ is 2.195 times larger, the transverse error is 46 times smaller for a given interval after its perihelion. The situation is summarized in Table \ref{tab:errors}.\\

After the reception of all spacecraft data, these 3I/ATLAS errors will probably go down by another factor of 5 or so. Projecting these uncertainties to a heliocentric distance of 100 AU reveals the stark difference in intercept feasibility. At this distance, the 1$\sigma$ positional errors are compared in Table \ref{tab:errors}.\\

\section{Conclusion}
Using 'Optimum Interplanetary Trajectory Software', the feasibility of exploiting a Solar Oberth Manoeuvre (SOM) as a method of catching the interstellar object 3I/ATLAS long past perihelion was studied. It was found this is in principle achievable with an optimal launch year of 2035 and utilising a Starship Block 3 upper stage fully-refuelled in Low Earth Orbit. Flight durations from 35 to 50 years are indicated, and masses around that of the New Horizons spacecraft (500 $\si{kg}$) are realisable. Challenges include more than one solid propellant stage would be needed for the SOM itself, a heat shield would have to be attached to protect the spacecraft from the high solar flux, and there would be a problematic sensitivity of the heliocentric escape asymptote to the SOM $\Delta$V.

\bibliography{SOto3I}{}

@ARTICLE{AH2,
       author = {{Hibberd}, Adam},
        title = "{Intermediate Points for Missions to Interstellar Objects Using Optimum Interplanetary Trajectory Software}",
      journal = {arXiv e-prints},
         year = 2022,
        month = may,
          eid = {arXiv:2205.10220},
        pages = {arXiv:2205.10220},
archivePrefix = {arXiv},
       eprint = {2205.10220},
 primaryClass = {astro-ph.EP},
       adsurl = {https://ui.adsabs.harvard.edu/abs/2022arXiv220510220H}
}

@misc{OITS_info,
title={Github repository for OITS.}, 
author={Adam Hibberd},
year={2017},
url={https://github.com/AdamHibberd/Optimum_Interplanetary_Trajectory}
}

@article{LeDigabel2011,
author = {{Le Digabel}, S.},
journal = {ACM Transactions on Mathematical Software (TOMS)},
number = {4},
pages = {44},
title = {{Algorithm 909: NOMAD: Nonlinear optimization with the MADS algorithm}},
volume = {37},
year = {2011}
}

@article{Schlueter_et_al_2013,
    author = {Schlueter, M. and Erb, S. and Gerdts, M. and Kemble, S. and Ruckmann, J.J.},
    title = {MIDACO on MINLP Space Applications},
    journal = {Advances in Space Research},
    publisher = {Elsevier},    
    volume = {51},
    number = {7},
    pages = {1116--1131},
    year = {2013},
    doi = {10.1016/j.asr.2012.11.006}
}

@article{Schlueter_et_al_2009,
    author = {Schlueter, M. and Egea, J.A. and Banga, J.R.},
    title = {Extended Ant Colony Optimization for non-convex Mixed Integer Nonlinear Programming},
    journal = {Computers and Operations Research},
    publisher = {Elsevier},    
    volume = {36},
    number = {7},
    pages = {2217--2229},
    year = {2009},
    doi = {10.1016/j.cor.2008.08.015}     
}

@article{Schlueter_Gerdts_2010,
    author = {Schlueter, M. and Gerdts, M.},
    title = {The Oracle Penalty Method},
    journal = {Journal of Global Optimization},
    publisher = {Springer},
    volume = {47},
    number = {2},
    pages = {293--325},
    year = {2010},
    doi = {10.1007/s10898-009-9477-0}    
}

@ARTICLE{HPE19,
       author = {{Hein}, Andreas M. and {Perakis}, Nikolaos and {Eubanks}, T. Marshall and {Hibberd}, Adam and {Crowl}, Adam and {Hayward}, Kieran and {Kennedy}, Robert G. and {Osborne}, Richard},
        title = "{Project Lyra: Sending a spacecraft to 1I/'Oumuamua (former A/2017 U1), the interstellar asteroid}",
      journal = {Acta Astronaut.},
         year = 2019,
        month = aug,
       volume = {161},
        pages = {552-561},
          doi = {10.1016/j.actaastro.2018.12.042},
       adsurl = {https://ui.adsabs.harvard.edu/abs/2019AcAau.161..552H}
}

@ARTICLE{HEL22,
       author = {{Hein}, Andreas M. and {Eubanks}, T. Marshall and {Lingam}, Manasvi and {Hibberd}, Adam and {Fries}, Dan and {Schneider}, Jean and {Kervella}, Pierre and {Kennedy}, Robert and {Perakis}, Nikolaos and {Dachwald}, Bernd},
        title = "{Interstellar Now! Missions to Explore Nearby Interstellar Objects}",
      journal = {Adv. Space Res.},
         year = 2022,
        month = jan,
       volume = {69},
       number = {1},
        pages = {402-414},
          doi = {10.1016/j.asr.2021.06.052},
       adsurl = {https://ui.adsabs.harvard.edu/abs/2022AdSpR..69..402H}
}

@ARTICLE{HHE20,
       author = {{Hibberd}, Adam and {Hein}, Andreas M. and {Eubanks}, T. Marshall},
        title = "{Project Lyra: Catching 1I/'Oumuamua - Mission opportunities after 2024}",
      journal = {Acta Astronautica},
         year = 2020,
        month = may,
       volume = {170},
        pages = {136-144},
          doi = {10.1016/j.actaastro.2020.01.018},
archivePrefix = {arXiv},
       eprint = {1902.04935},
 primaryClass = {physics.space-ph},
       adsurl = {https://ui.adsabs.harvard.edu/abs/2020AcAau.170..136H}
}

@ARTICLE{HH21,
       author = {{Hibberd}, Adam and {Hein}, Andreas M.},
        title = "{Project Lyra: Catching 1I/'Oumuamua-Using Nuclear Thermal Rockets}",
      journal = {Acta Astronaut.},
         year = 2021,
        month = feb,
       volume = {179},
        pages = {594-603},
          doi = {10.1016/j.actaastro.2020.11.038},
archivePrefix = {arXiv},
       eprint = {2008.05435},
 primaryClass = {astro-ph.IM},
       adsurl = {https://ui.adsabs.harvard.edu/abs/2021AcAau.179..594H}
}

@ARTICLE{HPH21,
       author = {{Hibberd}, Adam and {Perakis}, Nikolaos and {Hein}, Andreas M.},
        title = "{Sending a spacecraft to interstellar comet 2I/Borisov}",
      journal = {Acta Astronaut.},
         year = 2021,
        month = dec,
       volume = {189},
        pages = {584-592},
          doi = {10.1016/j.actaastro.2021.09.006},
       adsurl = {https://ui.adsabs.harvard.edu/abs/2021AcAau.189..584H}
}

@article{AH23,
  title="{Project Lyra: Another possible trajectory to 1I/’Oumuamua}",
  author={{Hibberd}, Adam},
  journal={Acta Astronaut.},
  volume={211},
  pages={431-434},
  year={2023},
    doi = {10.1016/j.actaastro.2023.06.029},
}

@ARTICLE{HA23,
       author = {{Hibberd}, Adam},
        title = "{Project Lyra: The Way to Go and the Launcher to Get There}",
      journal = {arXiv e-prints},
         year = 2023,
        month = may,
          eid = {arXiv:2305.03065},
        pages = {arXiv:2305.03065},
          doi = {10.48550/arXiv.2305.03065},
archivePrefix = {arXiv},
       eprint = {2305.03065},
 primaryClass = {astro-ph.IM},
       adsurl = {https://ui.adsabs.harvard.edu/abs/2023arXiv230503065H}
}

@misc{seligman2025discovery,
      title={Discovery and Preliminary Characterization of a Third Interstellar Object: 3I/ATLAS}, 
      author={Darryl Z. Seligman and Marco Micheli and Davide Farnocchia and Larry Denneau and John W. Noonan and Henry H. Hsieh and Toni Santana-Ros and John Tonry and Katie Auchettl and Luca Conversi and Maxime Devogèle and Laura Faggioli and Adina D. Feinstein and Marco Fenucci and Marin Ferrais and Tessa Frincke and Olivier R. Hainaut and Kyle Hart and Andrew Hoffman and Carrie E. Holt and Willem B. Hoogendam and Mark E. Huber and Emmanuel Jehin and Theodore Kareta and Jacqueline V. Keane and Michael S. P. Kelley and Tim Lister and Kathleen Mandt and Dušan Marčeta and Karen J. Meech and Mohamed Amine Miftah and Marvin Morgan and Francisco Ocaña and Eloy Peña-Asensio and Benjamin J. Shappee and Robert J. Siverd and Aster G. Taylor and Michael A. Tucker and Richard Wainscoat and Robert Weryk and James J. Wray and Atsuhiro Yaginuma and Bin Yang and Quanzhi Ye and Qicheng Zhang},
      year={2025},
      eprint={2507.02757},
      archivePrefix={arXiv},
      primaryClass={astro-ph.EP},
      url={https://arxiv.org/abs/2507.02757}, 
}

@misc{bolin2025,
      title={Interstellar comet 3I/ATLAS: discovery and physical description}, 
      author={Bryce T. Bolin and Matthew Belyakov and Christoffer Fremling and Matthew J. Graham and Ahmed. M. Abdelaziz and Eslam Elhosseiny and Candace L. Gray and Carl Ingebretsen and Gracyn Jewett and Sergey Karpov and Mukremin Kilic and Martin Mašek and Mona Molham and Diana Roderick and Ali Takey and Carey M. Lisse and Laura-May Abron and Michael W. Coughlin and Cheng-Han Hsieh and Keith S. Noll and Ian Wong},
      year={2025},
      eprint={2507.05252},
      archivePrefix={arXiv},
      primaryClass={astro-ph.EP},
      url={https://arxiv.org/abs/2507.05252}, 
}

@misc{opitom2025,
      title={Snapshot of a new interstellar comet: 3I/ATLAS has a red and featureless spectrum}, 
      author={Cyrielle Opitom and Colin Snodgrass and Emmanuel Jehin and Michele T. Bannister and Erica Bufanda and Sophie E. Deam and Rosemary Dorsey and Marin Ferrais and Said Hmiddouch and Matthew M. Knight and Rosita Kokotanekova and Brayden Leicester and Michaël Marsset and Brian Murphy and Vincent Okoth and Ryan Ridden-Harper and Mathieu Vander Donckt and Léa Ferellec and Damien Hutsemekers and Manuela Lippi and Jean Manfroid and Zouhair Benkhaldoun},
      year={2025},
      eprint={2507.05226},
      archivePrefix={arXiv},
      primaryClass={astro-ph.EP},
      url={https://arxiv.org/abs/2507.05226}, 
}

@misc{alvarezcandal2025,
      title={X-SHOOTER Spectrum of Comet C/2025 N1: Insights into a Distant Interstellar Visitor}, 
      author={A. Alvarez-Candal and J. L. Rizos and L. M. Lara and P. Santos-Sanz and P. J. Gutierrez and J. L. Ortiz and N. Morales},
      year={2025},
      eprint={2507.07312},
      archivePrefix={arXiv},
      primaryClass={astro-ph.EP},
      url={https://arxiv.org/abs/2507.07312}, 
}

@article{Loeb_2025,
doi = {10.3847/2515-5172/adee06},
url = {https://dx.doi.org/10.3847/2515-5172/adee06},
year = {2025},
month = {jul},
publisher = {The American Astronomical Society},
volume = {9},
number = {7},
pages = {178},
author = {Loeb, Abraham},
title = {3I/ATLAS is Smaller or Rarer than It Looks},
journal = {Research Notes of the AAS}
}

@misc{NAIF,
title="{Planetary Data System Navigation Node}",
author = "{NAIF}",
year = {2025},
url = {https://naif.jpl.nasa.gov/naif/aboutspice.html}
}

@BOOK{Bate1971,
       author = {{Bate}, R.~R. and {Mueller}, D.~D. and {White}, J.~E.},
        title = "{Fundamentals of astrodynamics}",
         year = 1971,
publisher = {New York: Dover Publications},
       adsurl = {https://ui.adsabs.harvard.edu/abs/1971fuas.book.....B}
}

@ARTICLE{Chandler2025,
       author = {{Chandler}, Colin Orion and {Bernardinelli}, Pedro H. and {Juri{\'c}}, Mario and {Singh}, Devanshi and {Hsieh}, Henry H. and {Sullivan}, Ian and {Jones}, R. Lynne and {Kurlander}, Jacob A. and {Vavilov}, Dmitrii and {Eggl}, Siegfried and {Holman}, Matthew and {Spoto}, Federica and {Schwamb}, Megan E. and {Christensen}, Eric J. and {Beebe}, Wilson and {Roodman}, Aaron and {Lim}, Kian-Tat and {Jenness}, Tim and {Bosch}, James and {Smart}, Brianna and {Bellm}, Eric and {MacBride}, Sean and {Rawls}, Meredith L. and {Greenstreet}, Sarah and {Slater}, Colin and {Heinze}, Aren and {Ivezi{\'c}}, {\v{Z}}eljko and {Blum}, Bob and {Connolly}, Andrew and {Daues}, Gregory and {Makadia}, Rahil and {Gower}, Michelle and {Bryce Kalmbach}, J. and {Monet}, David and {Bannister}, Michele T. and {Dones}, Luke and {Dorsey}, Rosemary C. and {Fraser}, Wesley C. and {Forbes}, John C. and {Fuentes}, Cesar and {Holt}, Carrie E. and {Inno}, Laura and {Jones}, Geraint H. and {Knight}, Matthew M. and {Lintott}, Chris J. and {Lister}, Tim and {Lupton}, Robert and {Mendoza Magbanua}, Mark Jesus and {Malhotra}, Renu and {Mueller}, Beatrice E.~A. and {Murtagh}, Joseph and {Pandey}, Nitya and {Reach}, William T. and {Samarasinha}, Nalin H. and {Seligman}, Darryl Z. and {Snodgrass}, Colin and {Solontoi}, Michael and {Szab{\'o}}, Gyula M. and {White}, Ellie and {Womack}, Maria and {Young}, Leslie A. and {Allbery}, Russ and {Armellin}, Roberto and {Aubourg}, {\'E}ric and {Avdellidou}, Chrysa and {Azfar}, Farrukh and {Bauer}, James and {Bechtol}, Keith and {Belyakov}, Matthew and {Benecchi}, Susan D. and {Bertini}, Ivano and {Bolin}, Bryce T. and {Bose}, vMaitrayee and {Buchanan}, Laura E. and {Boucaud}, Alexandre and {Boufleur}, Rodrigo C. and {Boutigny}, Dominique and {Braga-Ribas}, Felipe and {Calabrese}, Daniel and {Camargo}, J.~I.~B. and {Caplar}, Neven and {Carry}, Benoit and {Carvajal}, Juan Pablo and {Choi}, Yumi and {Cowan}, Preeti and {Croft}, Steve and {{\'C}uk}, Matija and {Daruich}, Felipe and {Daubard}, Guillaume and {Davenport}, James R.~A. and {Daylan}, Tansu and {Delgado}, Jennifer and {Devillepoix}, Hadrien A.~R. and {Doherty}, Peter E. and {Donaldson}, Abbie and {Drass}, Holger and {Deppe}, Stephanie JH and {Dubois-Felsmann}, Gregory P. and {Economou}, Frossie and {Eduardo}, Marielle R. and {Farnocchia}, Davide and {Frissell}, Maxwell K. and {Fedorets}, Grigori and {Fernandes}, Maryann Benny and {Fulle}, Marco and {Gerdes}, David W. and {Gibbs}, Alex R. and {Gillan}, A. Fraser and {Guy}, Leanne P. and {Hammergren}, Mark and {Hanushevsky}, Andrew and {Hernandez}, Fabio and {Hestroffer}, Daniel and {Hopkins}, Matthew J. and {Granvik}, Mikael and {Ieva}, Simone and {Irving}, David H. and {Jannuzi}, Buell T. and {Jimenez}, David and {Ramos Gomes-J{\'u}nior}, Altair and {Juramy}, Claire and {Kahn}, Steven M. and {Kannawadi}, Arun and {Kang}, Yijung and {Kryszczy{\'n}ska}, Agnieszka and {Kotov}, Ivan and {Koumjian}, Alec and {Krughoff}, K. Simon and {Lage}, Craig and {Lange}, Travis J. and {Levine}, W. Garrett and {Li}, Zhuofu and {Licandro}, Javier and {Lin}, Hsing Wen and {Lust}, Nate B. and {Lyttle}, Ryan R. and {Mahabal}, Ashish A. and {Mahlke}, Max and {Plazas Malag{\'o}n}, Andr{\'e}s A. and {Salazar Manzano}, Luis E. and {Marc}, Moniez and {Margoti}, Giuliano and {Mar{\v{c}}eta}, Du{\v{s}}an and {Menanteau}, Felipe and {Meyers}, Joshua and {Mills}, Dave and {Morato}, Naomi and {More}, Surhud and {Morrison}, Christopher B. and {Moulane}, Youssef and {Mu{\~n}oz-Guti{\'e}rrez}, Marco A. and {Newcomer F.}, M. and {O'Connor}, Paul and {Oldag}, Drew and {Oldroyd}, William J and {O'Mullane}, William and {Opitom}, Cyrielle and {Oszkiewicz}, Dagmara and {Page}, Gary L. and {Patterson}, Jack and {Payne}, Matthew J. and {Peloton}, Julien and {Pereira}, Chrystian Luciano and {Peterson}, John R. and {Polin}, Daniel and {Pollek}, Hannah Mary Margaret and {Polen}, Rebekah and {Qiu}, Yongqiang and {Ragozzine}, Darin and {Rajagopal}, Jayadev and {van Reeven}, vWouter and {Rice}, Malena and {Ridgway}, Stephen T. and {Rivkin}, Andrew S. and {Robinson}, James E. and {Ro{\.z}ek}, Agata and {Salnikov}, Andrei and {S{\'a}nchez}, Bruno O. and {Sarid}, Gal and {Schambeau}, Charles A. and {Scolnic}, Daniel and {Schindler}, Rafe H. and {Seaman}, Robert and {Jacques}, {\v{S}}ebag and {Shaw}, Richard A. and {Shugart}, Alysha and {Sick}, Jonathan and {Siraj}, Amir and {Sitarz}, Michael C. and {Sobhani}, Shahram and {Soldahl}, Christine and {Stalder}, Brian and {Stetzler}, Steven and {Swinbank}, John D. and {Szigeti}, L{\'a}szl{\'o} and {Tauraso}, Michael and {Thornton}, Adam and {Tonietti}, Luca and {Trilling}, David E. and {Trujillo}, Chadwick A.},
        title = "{NSF-DOE Vera C. Rubin Observatory Observations of Interstellar Comet 3I/ATLAS (C/2025 N1)}",
      journal = {arXiv e-prints},
         year = 2025,
        month = jul,
          eid = {arXiv:2507.13409},
        pages = {arXiv:2507.13409},
          doi = {10.48550/arXiv.2507.13409},
archivePrefix = {arXiv},
       eprint = {2507.13409},
 primaryClass = {astro-ph.EP},
       adsurl = {https://ui.adsabs.harvard.edu/abs/2025arXiv250713409C}
}

@ARTICLE{Belyakov2025,
       author = {{Belyakov}, Matthew and {Fremling}, Christoffer and {Graham}, Matthew J. and {Bolin}, Bryce T. and {Kilic}, Mukremin and {Jewett}, Gracyn and {Lisse}, Carey M. and {Ingebretsen}, Carl and {Davis}, M. Ryleigh and {Wong}, Ian},
        title = "{Palomar and Apache Point Spectrophotometry of Interstellar Comet 3I/ATLAS}",
      journal = {Research Notes of the American Astronomical Society},
         year = 2025,
        month = jul,
       volume = {9},
       number = {7},
          eid = {194},
        pages = {194},
          doi = {10.3847/2515-5172/adf059},
       adsurl = {https://ui.adsabs.harvard.edu/abs/2025RNAAS...9..194B}
}

@article{Yaginuma_2025,
doi = {10.3847/1538-4357/ae11b2},
url = {https://doi.org/10.3847/1538-4357/ae11b2},
year = {2025},
month = {dec},
publisher = {The American Astronomical Society},
volume = {995},
number = {1},
pages = {64},
author = {Yaginuma, Atsuhiro and Frincke, Tessa and Seligman, Darryl Z. and Mandt, Kathleen and DellaGiustina, Daniella N. and Peña-Asensio, Eloy and Taylor, Aster G. and Nolan, Michael C.},
title = {The Feasibility of a Spacecraft Flyby with the Third Interstellar Object 3I/ATLAS from Earth or Mars},
journal = {The Astrophysical Journal}
}

@article{loeb2025intercepting,
  title={Intercepting 3I/ATLAS at Its Closest Approach to Jupiter with the Juno Spacecraft},
  author={Loeb, Abraham and Hibberd, Adam and Crowl, Adam},
  journal={Aerospace},
  volume={12},
  number={9},
  pages={851},
  year={2025},
  publisher={MDPI}
}

@article{sanchez2025analysis,
  title={Analysis of Trajectories to 3I/ATLAS with a Comet Interceptor-like Spacecraft},
  author={Sanchez, Joan Pau and Snodgrass, Colin},
  journal={Research Notes of the AAS},
  volume={9},
  number={7},
  pages={207},
  year={2025},
  publisher={The American Astronomical Society}
}

@book{oberth2019wege,
  title={Wege zur raumschiffahrt},
  author={Oberth, Hermann},
  year={2019},
  publisher={Walter de Gruyter GmbH \& Co KG}
}

@misc{I4IS,
title="{Initiative for Interstellar Studies}",
author = "{i4is}",
year = {2025},
url = {https://i4is.org/}
}

@article{Seligman_2018,
doi = {10.3847/1538-3881/aabd37},
url = {https://dx.doi.org/10.3847/1538-3881/aabd37},
year = {2018},
month = {apr},
publisher = {The American Astronomical Society},
volume = {155},
number = {5},
pages = {217},
author = {Seligman, Darryl and Laughlin, Gregory},
title = {The Feasibility and Benefits of In Situ Exploration of ‘Oumuamua-like Objects},
journal = {The Astronomical Journal}
}

@article{hibberd2025interstellar,
  title={Is the Interstellar Object 3I/ATLAS Alien Technology?},
  author={Hibberd, Adam and Crowl, Adam and Loeb, Abraham},
  journal={arXiv preprint arXiv:2507.12213},
  year={2025}
}

@inproceedings{benkoski2019physics,
  title={The physics of heat shielding during an oberth maneuver},
  author={Benkoski, Jason J and McNutt, Ralph and Brandt, Pontus and Paul, Michael V},
  booktitle={Proceedings of the International Astronautical Congress, IAC},
  volume={2019},
  pages={IAC--19\_D4\_4\_4\_x51096},
  year={2019}
}

@article{hein2022interstellar,
  title={Interstellar now! Missions to explore nearby interstellar objects},
  author={Hein, Andreas M and Eubanks, T Marshall and Lingam, Manasvi and Hibberd, Adam and Fries, Dan and Schneider, Jean and Kervella, Pierre and Kennedy, Robert and Perakis, Nikolaos and Dachwald, Bernd},
  journal={Advances in Space Research},
  volume={69},
  number={1},
  pages={402--414},
  year={2022},
  publisher={Elsevier}
}
\bibliographystyle{aasjournalv7}
\end{document}